\documentclass[letterpaper, 10 pt, conference]{ieeeconf}  

\IEEEoverridecommandlockouts                              
\usepackage{xcolor,amsmath,amssymb,mathtools,booktabs,tikz,pgfplots}
\DeclareMathOperator*{\minimize}{minimize}
\DeclareMathOperator*{\subject_to}{s. t.\: }
\newtheorem{theorem}{Theorem}
\newtheorem{assumption}{Assumption}
\newtheorem{definition}{Definition}
\newtheorem{remark}{Remark}
\newtheorem{corollary}{Corollary}
\pgfplotsset{every axis/.append style={
		label style={font=\large},
		tick label style={font=\large}  
}}

\title{\LARGE \bf
Robust Stability of Gaussian Process Based Moving Horizon Estimation
}

\author{Tobias M. Wolff, Victor G. Lopez, and Matthias A. Müller 
\thanks{This project has received funding from the European Research
	Council (ERC) under the European Union’s Horizon 2020 research
	and innovation programme (grant agreement No 948679).}
\thanks{Tobias M. Wolff, Victor G. Lopez, and Matthias A. Müller are with the Leibniz University Hannover, Institute of Automatic Control, Germany
        {\tt\small \{wolff,lopez,mueller\}@irt.uni-hannover.de}}%
}

\usepackage{fancyhdr}

\fancyfootoffset{-1em}
\fancypagestyle{copyright}{
	\chead{\footnotesize This work has been submitted to the IEEE for possible publication. Copyright may be transferred without notice, after
		which this version may no longer be accessible.}
}

\begin{document}

\maketitle
\thispagestyle{copyright}

\begin{abstract}
In this paper, we introduce a Gaussian process based moving horizon estimation (MHE) framework. The scheme is based on offline collected data and offline hyperparameter optimization. In particular, compared to standard MHE schemes, we replace the mathematical model of the system by the posterior mean of the Gaussian process. To account for the uncertainty of the learned model, we exploit the posterior variance of the learned Gaussian process in the weighting matrices of the cost function of the proposed MHE scheme. We prove practical robust exponential stability of the resulting estimator using a recently proposed Lyapunov-based proof technique. Finally, the performance of the Gaussian process based MHE scheme is illustrated via a nonlinear system.
\end{abstract}

\section{INTRODUCTION}
Moving horizon estimation (MHE) \cite{Rawlings2020,Rao2001,Alessandri2005,Gharbi2019} is a nonlinear, optimization-based state estimation technique. Loosely speaking, at each time instant, we first measure the current output of the system. Then, we solve an optimization problem to determine an optimal estimated state sequence over some (finite) estimation horizon. Inherent physical constraints of the system, such as, e.g., nonnegativity constraints of chemical concentrations or hormone concentrations can be accounted for in the optimization problem. Finally, the state estimate is set to the last element of the optimal estimated state sequence. MHE is particularly suitable for nonlinear state estimation, as it can outperform other nonlinear state estimation techniques such as the extended Kalman filter \cite{Rawlings2020}.

However, MHE crucially relies on the knowledge of an accurate mathematical model of the dynamical system. The derivation of such a mathematical model from first principles can be difficult, time-consuming, and expensive. Alternatively, an MHE scheme can be set up by solely relying on data, or by learning the system dynamics using some machine learning technique. In this work, we focus on the latter approach, namely by learning a mathematical model of the system dynamics using Gaussian Processes (GPs) \cite{Williams2006}. 

GPs are a Bayesian machine learning technique which are defined as a collection of random variables, any finite number of which follows a joint Gaussian distribution \cite[Def. 2.1]{Williams2006}. In recent years, GPs have been increasingly used in the area of learning-based control (compare, e.g., \cite{Beckers2019,Hewing2019,Maiworm2021}). Here, an advantage is that they inherently allow for a quantification of the model uncertainty, which is typically not the case when using other machine learning techniques such as, e.g., neural networks. 

Concerning the design of learning-based estimators, there exist only few results in the literature, both in the context of MHE and regarding the usage of GP-based techniques. In \cite{Alessandri2011}, a so-called state estimation function is learned by means of a feedforward neural network. The authors in \cite{Cao2022} develop a learning-based MHE scheme, where the mapping of the input data (including the system matrices and the measured outputs) to the state estimates is learned offline. Furthermore, GPs have been used to estimate time-delays within a standard MHE scheme \cite{Mori2020}. Moreover, GPs have been exploited in the design of an extended Kalman Filter \cite{Ko2009} and to develop a joint dynamics and state estimation framework \cite{Buisson2021}.

The contribution of this work is the introduction of a GP based nonlinear MHE framework. We exploit the posterior mean of the GP to approximate the system dynamics and the posterior variance in the design of the weighting matrices of the MHE. The advantages of this approach are two-fold. First, we do not require that any mathematical model of the system dynamics is available a priori. Second, we directly account for the uncertainty of the learned model by using weighting matrices that depend on the regression\footnote{In this paper, we clearly distinguish between \textit{regression} inputs to the GP and \textit{control} inputs to the physical system.} inputs. Furthermore, we prove practical robust exponential stability of the resulting estimator. 
In a numerical example, we illustrate the performance of the proposed GP based MHE scheme. 

The outline of this paper is the following. In Section~\ref{sec:pre}, we revise the fundamentals of GPs and introduce the setting of our work. Then, we introduce the GP based MHE scheme and show the related robust stability proof in Sections~\ref{sec:GP-MHE} and~\ref{sec:main:stability}, respectively. Finally, we close this paper with a numerical example and a conclusion in Sections~\ref{sec:Numerical-Example} and~\ref{sec:Conclusion}, respectively.

\section{PRELIMINARIES AND SETTING}
\label{sec:pre}
We denote the set of integers greater than or equal to $a \in \mathbb{R}$ by $\mathbb{I}_{\geq a}$. The set of non-negative real numbers is denoted by $\mathbb{R}_{\geq 0}$. The weighted vector norm for a vector $x = [x_1 \dots x_n]^\top \in \mathbb{R}^n$ and a symmetric positive definite matrix~$P$ is written as $\|x\|_P = \sqrt{x^\top P x} $. The identity matrix of dimension~$n$ is denoted by $I_n$. A diagonal matrix of dimension $n$ with $q_1, \dots, q_n$ on the diagonal entries is written as $\mathrm{diag}(q_1, \dots, q_n)$. The standard maximal eigenvalue of a positive definite matrix~$P$ is denoted by $\lambda_{\max}(P)$. The maximum generalized eigenvalue of square matrices $P_1$ and $P_2$ is denoted by $\lambda_{\max}(P_1,P_2)$. For two symmetric matrices~$A,~B,~A\leq B$ means that $(B-A)$ is positive semidefinite.

GPs are an increasingly popular method to approximate a nonlinear function $f(d)$. GPs are fully defined by a mean function $m: \mathcal{Z} \rightarrow \mathbb{R}$ (where $\mathcal{Z} \subseteq \mathbb{R}^{n_d}$) and a covariance function (also referred to as kernel) $k: \mathcal{Z} \times \mathcal{Z} \rightarrow \mathbb{R}$ for some regression inputs $d$, $d' \in \mathcal{Z}$
\begin{align}
	f(d) \sim \mathcal{GP}(m(d), k(d,d')).
\end{align}
In this work, we consider (as commonly done in the context of GPs) a prior mean $m \equiv 0$ and the squared exponential automatic relevance determination (ARD) kernel, i.e., 
\begin{align*}
	k(d_i,d_j) \coloneqq \sigma_f^2 \exp (-\frac{1}{2}(d_i-d_j)^\top \Lambda^{-1}(d_i-d_j)) + \sigma_\varepsilon^2\delta_{ij}, 
\end{align*}
where $\sigma_f \in \mathbb{R}_{\geq 0}$, $\Lambda = \mathrm{diag}(\varphi_1^2, \varphi_2^2, \dots, \varphi_{n_d}^2)$ with~$\varphi_1,~\varphi_2,~\dots,~\varphi_{n_d}  \in \mathbb{R}_{\geq 0}$, and $\sigma_\varepsilon \in \mathbb{R}_{\geq 0}$. The symbol~$\delta_{ij}$ denotes the Kronecker delta.

The GP is trained by conditioning it on some given regression input data\footnote{Here, we use $D^d$ to denote the regression input data (instead of the commonly used notation~$X$) to avoid confusion with respect to the actual system states.} $D^d =\begin{bmatrix} d^d_1 & d^d_2 & \dots & d^d_N\end{bmatrix}$ and some output data $Y^d =\begin{bmatrix} y^d_1& y^d_2 & \dots & y^d_N\end{bmatrix}$, where each output data point $y^d$ is given by $f(d^d)+\varepsilon^d$ with $\varepsilon^d$ being normally distributed with distribution $\mathcal{N}(0,\sigma_\varepsilon^2)$. Then, the posterior mean at some regression test input $d_*$ is given by 
\begin{align}
	\label{post_mean}
	m_+(d_\ast|D^d,Y^d) = k(d_\ast,D^d)(K(D^d,D^d) + \sigma_\varepsilon^2I)^{-1}Y^d
\end{align}
and the posterior variance (which corresponds to the inherent uncertainty quantification) by
\begin{align}
	\label{post_var}
	\sigma_+^2&(d_\ast|D^d,Y^d) = \nonumber \\
	 &k(d_\ast,d_\ast) - k(d_\ast,D^d) (K(D^d,D^d) + \sigma_\varepsilon^2I)^{-1}k(D^d,d_\ast),
\end{align}
where $k(d_\ast,D^d) = \begin{pmatrix}
	k(d_\ast,d_i)
\end{pmatrix}_{d_i \in D^d} = k(D^d,d_\ast)^\top$, with~$k(d_\ast,D^d) \in \mathbb{R}^{1 \times N}$, and $K(D^d,D^d) = (k(d_i,d_j))_{d_i, d_j \in D^d}$ with~$K(D^d,D^d) \in \mathbb{R}^{N \times N}$. The posterior mean depends on the hyperparameters~$\sigma_f, \varphi_1, \dots, \varphi_{n_d}, \sigma_\varepsilon$ that crucially influence the quality of the learned model, compare the discussion in \cite[Sec. 2.3]{Williams2006}. As commonly done in the literature, we determine the hyperparameters by maximizing the log marginal likelihood, see, e.g., \cite[Eq. (2.30)]{Williams2006}.

After this general introduction to GPs, we now describe how they are exploited in this work. We consider discrete-time nonlinear systems with additive disturbances, i.e., 
\begin{subequations}
	\label{def:system}
	\begin{align}
	x(t+1) &= f(x(t),u(t)) + w(t) \\
	y(t) &= h(x(t), u(t)) + v(t)
	\end{align}
\end{subequations}
with $x(t), w(t) \in \mathbb{R}^n$, $u(t) \in \mathbb{R}^m$, and $y(t), v(t) \in \mathbb{R}^p$, where $w$ and $v$ denote the process and the measurement noise, respectively.
	
Throughout this paper, we assume that the states and inputs evolve in compact sets, i.e., $x(t) \in \mathbb{X} \subset \mathbb{R}^n$ and~$u(t) \in \mathbb{U} \subset \mathbb{R}^m \: \: \forall t \in \mathbb{I}_{\geq 0}$. In the following, we model~$f$ and~$h$ using GPs. Hence, for modeling purposes only, we consider~$w$ and~$v$ to be normally distributed. Note that such a setting (assuming bounded states and employing GP models) is common in the GP-based control/estimation literature, compare, e.g., \cite{Maiworm2021,Buisson2021}. This corresponds to the realistic scenario in which the real process and measurement disturbances~$w$ and~$v$ in (\ref{def:system}) are not unbounded in practical applications, despite being assumed to be normally distributed within the GP modeling.
The hyperparameter~$\sigma_\varepsilon$ is determined such that the GP approximates the unknown function as well as possible, compare the discussion in \cite[Sec. II C]{Buisson2021}. Furthermore, we assume that the state transition function~$f$ as well as the output mapping~$h$ are continuous. 

The objective is to learn the state-space model~(\ref{def:system}) (i.e., the state transition function $f$ and the output mapping $h$) by means of GPs. In this case, the regression input data is composed of the states and the control inputs at time~$t$, i.e., $d(t) = \begin{bmatrix}
	x_1(t) & \dots & x_n(t) & u_1(t) & \dots & u_m(t) 
\end{bmatrix}^\top$. Since standard GPs only map on scalar regression outputs, we need to learn $n + p$ independent GPs to approximate the complete dynamics (i.e., $n$ GPs for the components $f_{1}$,~$f_{2}$, \dots,~$f_{n}$ of the function $f$, and $p$ GPs for the components $h_{1},~h_{2},\dots, ~h_{p}$ of the function $h$). In the following, we denote the kernels associated to the GPs approximating the components of~$f$ and~$h$ by $k_{x_1},~k_{x_2}, \dots, k_{x_n}$ and $k_{y_1},~k_{y_2}, \dots, k_{y_p}$, respectively. We collect training data, condition the GPs on the training data, and tune the hyperparameters by maximizing the marginal log-likelihood. To simplify the notation, we write
\begin{align}
	m_{+,x} = \begin{pmatrix}
		m_{+, x_1} \\
		m_{+, x_2} \\
		\vdots \\
		m_{+, x_n} 
	\end{pmatrix} \quad
	m_{+,y} = \begin{pmatrix}
		m_{+, y_1} \\
		m_{+, y_2} \\
		\vdots \\
		m_{+, y_p} 
	\end{pmatrix},
\end{align}
which denote the vectors of the posterior means and 
\begin{align}
	\sigma^{2}_{+,x} =\begin{pmatrix}
		\sigma_{+,x_1}^{2} \\
		\sigma_{+,x_2}^{2} \\
		\vdots \\
		\sigma_{+,x_n}^{2}
	\end{pmatrix} \quad
	\sigma^{2}_{+,y} = \begin{pmatrix}
		\sigma_{+,y_1}^{2} \\
		\sigma_{+,y_2}^{2} \\
		\vdots \\
		\sigma_{+,y_p}^{2}
	\end{pmatrix},
\end{align}
which denote the vectors of the posterior variances. Hence, the learned system dynamics can be expressed as 
\begin{subequations}
	\label{learned_system}
	\begin{align}
		x(t+1) &= m_{+,x}(d(t)|D^d,Y^d) + \check{w}(t) \\
		y(t) &= m_{+,y} (d(t)|D^d, Y^d) + \check{v}(t)
	\end{align}
\end{subequations} 
with $\check{w} \in \mathbb{R}^n$ and $\check{v} \in \mathbb{R}^p$. Note that we recover the original system (\ref{def:system}) for 
\begin{align}
	\check{w}(t) & \coloneqq f\big(x(t), u(t)\big) - m_{+,x}\big(d(t)|D^d,Y^d\big) + w(t) \label{def:auxiliary_w}\\
	\check{v}(t) & \coloneqq h\big(x(t), u(t)\big) - m_{+,y}\big(d(t)|D^d,Y^d\big) + v(t) \label{def:auxiliary_v}. 
\end{align}
In this work, we consider two different phases. On the one hand, an offline phase, in which noise-free measurements of the control inputs and noisy measurements of the outputs \emph{and} states are available. This assumption allows us to condition the GPs on the training data and to perform the hyperparameter optimization offline. On the other hand, an online phase, in which noise-free measurements of the control inputs, but only noisy measurements of the outputs (but \emph{not} the states) are available, meaning that the states must be estimated. To perform the state estimation, we apply the GP based MHE, which is explained in the following section. The assumption of having noisy state measurements available in an offline phase might be restrictive in general, but is certainly fulfilled in cases where one can measure all the states in a laboratory setting using sophisticated hardware that is not available online, compare also the discussion in \cite{Turan2021,Wolff2022}.
 
\section{GP BASED MHE SCHEME}
\label{sec:GP-MHE}
In this section, we explain in detail the GP based MHE scheme. As usual in MHE, at each time step $t$, an optimization problem is solved taking the past measurements over some horizon $M$ into account. Namely, 
\begin{subequations} \label{MHE_nom}
	\begin{align}
		\minimize_{\substack{\bar{x}(t- M_t |t), \bar{w}(\cdot|t)}} \hspace{0.2cm}  &J\big(\bar{x}(t-M_t|t), \bar{w}(\cdot|t), \bar{v}(\cdot|t),t\big) \label{cost_function_nom} \\
		\subject_to  \bar{x}(j+1|t) =&~m_{+,x}(\bar{d}(j|t)|D^d,Y^d) +\bar{w}(j|t), \label{eq:mean_in_MHE_f}\\
		y(j) =&~m_{+,y}(\bar{d}(j|t)|D^d,Y^d) + \bar{v}(j|t), \label{eq:mean_in_MHE_h}\\ 
		&\hspace{2.5cm}\forall j \in \mathbb{I}_{[t-M_t, t-1]} \nonumber \\
		\bar{x}(j|t) &\in \mathbb{X} \quad \forall j \in \mathbb{I}_{[t-M_t, t]}
	\end{align}
	with $M_t = \min\{t, M\}$ ($M$ being the horizon length),  
	\begin{align*}
		\bar{d}(j|t) \coloneqq \begin{bmatrix}
			\bar{x}_1(j|t) &
			\dots &
			\bar{x}_n(j|t) &
			u_1(j) &
			\dots &
			u_m(j) 
		\end{bmatrix}^\top
	\end{align*}
	and
	\begin{align}
		J(\bar{x}(t-&M_t|t),\bar{w}(\cdot|t), \bar{v}(\cdot|t),t)  \nonumber\\
		 \coloneqq& 2||\bar{x}(t-M_t|t) - \hat{x}(t-M_t)||_{P_2}^2 \eta^{M_t}\nonumber \\
		&+ \sum_{j = 1}^{M_t} 2\eta^{j-1}\Big( ||\bar{w}(t-j|t)||_{Q_{\bar{d}(t-j|t)}^{-1}}^2 \nonumber \\
		&\qquad \qquad \qquad + ||\bar{v}(t-j|t)||_{R_{\bar{d}(t-j|t)}^{-1}}^2\Big). \label{eq:cost:function}
		\end{align}
\end{subequations}
In (\ref{eq:cost:function}), $\eta \in [0,1)$ is a discount factor. The notation $\bar{d}(j|t)$ denotes the estimated state (together with the measured control input) at time~$j$, estimated at time~$t$. The estimated process and measurement noise trajectories, estimated at time~$t$, are denoted by $\bar{w}(\cdot|t)$ and $\bar{v}(\cdot|t)$, respectively. The cost function is composed of two terms: the prior weighting and the stage cost. Hence, the cost function trades off how much we believe the measurements within the current horizon and how much we believe the prior $\hat{x}(t-M_t)$. The optimal estimated state sequence is denoted by $\hat{x}(\cdot|t)$ (analogous for $\hat{d}(\cdot|t)$) and the estimated system state at time $t$ is set to the last element of the estimated state sequence, i.e., $\hat{x}(t) \coloneqq \hat{x}(t|t)$. 

Note the first main difference to standard model-based MHE schemes in (\ref{eq:mean_in_MHE_f}) and in (\ref{eq:mean_in_MHE_h}). We exploit the posterior mean functions of the GPs to approximate the state transition function~$f$ and the output mapping~$h$. 

The weighting matrices in the cost function are chosen as
\begin{align}
	Q_{\bar{d}(t-j|t)} =&~\mathrm{diag}\big(\sigma_{+,x_1}^{2}(\bar{d}(t-j|t)|D^d,Y^d), \dots, \nonumber \\ & \hspace{1cm} \sigma_{+,x_n}^{2}(\bar{d}(t-j|t)|D^d,Y^d)\big) + Q_0\\
	R_{\bar{d}(t-j|t)} =&~ \mathrm{diag}\big(\sigma_{+,y_1}^{2}(\bar{d}(t-j|t)|D^d,Y^d), \dots, \nonumber \\ & \hspace{1cm} \sigma_{+,y_p}^{2}(\bar{d}(t-j|t)|D^d,Y^d) \big) + R_0
\end{align}
with $Q_0$, $R_0$ positive definite (and $P_2$ positive definite). This choice of the weighting matrices constitutes the second main difference to standard MHE schemes. The weighting matrices $Q_{\bar{d}(t-j|t)}$ and $R_{\bar{d}(t-j|t)}$ are a sum of two matrices. The first one is a diagonal matrix, where the diagonal entries correspond to the posterior variances of the corresponding states/outputs, as in the work related to GP based extended Kalman filtering \cite{Ko2009}.
Loosely speaking, the beneficial effect of this choice is the following: in a region of low training data availability, the posterior variance, representing the uncertainty of the learned model, is rather high. In turn, the inverse weighing matrices, on which the cost function is based, induce a low weight on $\bar{w}$ and $\bar{v}$. Consequently, we allow for large magnitudes of $\bar{w}$ and $\bar{v}$. This is meaningful, since in areas of low training data availability, the mean functions will be poor approximations of the true functions~$f$ and~$h$. In the opposite situation, namely in areas of high training data availability, the posterior variance will be rather small. This leads to a high weight in the cost function, meaning that we only allow for small $\bar{w}$ and $\bar{v}$. Once again, this is a meaningful conclusion, since the learned model will be of high quality in such areas. The second matrix corresponds to the standard MHE weighting matrix. The matrices $Q_0$ and $R_0$ are typically set according to the variance of the process/measurement noise affecting the online measurements \cite{Rao2000}. The choice of the matrix $P_2$ is more difficult in the general nonlinear case \cite[Sec. 3.1]{Rao2000}. One model-based approach to design this matrix has recently been proposed in \cite[Cor. 3]{Schiller2022}.

The proposed MHE scheme is in the so-called prediction form, since the state at time $t$ is ``predicted"\footnote{The alternative is the ``filtering" form of MHE, where the output measurement at time $t$ is used to estimate the state at time $t$. The prediction form of MHE allows for an easier theoretical analysis, while the filtering form of MHE typically performs better in practice, compare the discussion in \cite{Allan2019}. We conjecture that all theoretical results presented here also hold for the filtering form of MHE.} based on the output measurements up to time $t-1$. As long as~$t < M$, we use the so-called full information estimator, i.e., $M_t = t$, meaning that all available measurements are taken into account.
\section{ROBUST STABILITY ANALYSIS}
\label{sec:main:stability}
In this section, we prove robust stability of the GP based MHE scheme based on the following definition, which is similar to \cite[Def. 2]{Schiller2022}, with the main difference that we consider a \emph{practical} stability notion that can capture the mismatch between the posterior means and the true functions~$f$ and~$h$. 
\begin{definition}
	\label{def:pRES}
	A state estimator for system~(\ref{def:system}) is practically robustly exponentially stable (pRES) if there exist $C_1, C_2, C_3 > 0$, $\lambda_1, \lambda_2,\lambda_3 \in [0,1)$, and $\alpha >0$ such that the resulting state estimates $\hat{x}(t)$ satisfy
	\begin{align}	
		\label{eq:pRES}	
		\|x&(t)-\hat{x}(t)\| \leq \max \Big\{ C_1\|x(0)-\hat{x}(0)\| \lambda_1^t, \nonumber \\ &\max_{j\in\mathbb{I}_{[0,t-1]}}C_{2}\|w(j)\| \lambda_2^{t-j-1},\max_{j\in\mathbb{I}_{[0,t-1]}}C_{3}\|v(j)\| \lambda_3^{t-j-1}, \alpha \Big\}
	\end{align}
	for all $t\in\mathbb{I}_{\geq0}$, all initial conditions $x(0),\hat{x}(0) \in \mathbb{X}$, and every trajectory $(x(t),u(t),w(t),v(t))_{t=0}^\infty$ satisfying the system dynamics~(\ref{def:system}).
\end{definition}

Next, we introduce the matrices $Q_{\min}^{-1},~R_{\min}^{-1}$ and $Q_{\max}^{-1},~R_{\max}^{-1}$ such that 
\begin{align}
	Q_{\min}^{-1}  &\leq Q^{-1}_{\bar{d}(t)} \leq Q_{\max}^{-1}, \label{eq:Q} \\
	R_{\min}^{-1}  &\leq R^{-1}_{\bar{d}(t)} \leq R_{\max}^{-1}. \label{eq:R}
\end{align}
The matrices $Q_{\min}^{-1}$ and~$R_{\min}^{-1}$ represent the case when the regression test inputs are (infinitely) far away from the regression training inputs, meaning that the posterior variance is maximal. In case of the here considered squared exponential ARD kernel, an upper bound for the maximal posterior variance is given by $\sigma_f^2 + \sigma_\varepsilon^2$, i.e., a lower bound for $Q_{\min}^{-1}$ is given by
\begin{align}
	Q_{\min}^{-1} \coloneqq \big[ \mathrm{diag}(\sigma_{f, x_1}^2 + \sigma_{\varepsilon,x_1}^2, \dots, \sigma_{f, x_n}^2+ \sigma_{\varepsilon,x_n}^2) + Q_0\big]^{-1},
\end{align}
and $R_{\min}^{-1}$ is defined analogously. In turn, the matrices~$Q_{\max}^{-1},~R_{\max}^{-1}$ correspond to the minimal possible posterior variances. A lower bound for the minimal posterior variance is 0, which occurs in the noise-free case, when a regression test input corresponds exactly to a regression training input. Consequently, an upper bound for $Q_{\max}^{-1}$ (and similarly $R_{\max}^{-1}$) is given by
\begin{align}
	Q_{\max}^{-1} \coloneqq Q_0^{-1}.
\end{align}

To prove robust stability of the GP based MHE scheme, we need a detectability assumption. The assumption applied in this work is proposed in a similar way in \cite{Schiller2022} with the main difference that we here consider the learned system dynamics and not the true system dynamics. 
\begin{assumption}
	\label{ass:lyap}
	The system~(\ref{learned_system}) admits a $\delta$-IOSS Lyapunov function $W_\delta: \mathbb{R}^n \times \mathbb{R}^n \rightarrow \mathbb{R}_{\geq 0}$ with quadratic bounds and supply rate, i.e., there exist $ \eta\in[0,1),~P_1,~P_2,~Q_0,~R_0>0$ such that
	\begin{subequations}
		\begin{align}
			\label{ass:lyap:bounds}
			||x - \tilde{x}||_{P_{1}}^2 \leq W_\delta(x,\tilde{x}) \leq ||x-\tilde{x}||_{P_{2}}^2, 
		\end{align}
		\begin{align}
			\label{ass:lyap:supply}
			W_\delta \Big(&m_{+,x}(d|D^d,Y^d) + \check{w}, m_{+,x}(\tilde{d}|D^d,Y^d) + \check{w}^\prime\Big) \nonumber \\
			&\leq \eta W_\delta(x, \tilde{x}) + ||\check{w} - \check{w}^\prime||_{Q_{\min}^{-1}}^2  \nonumber \\  &+||m_{+,y}(d|D^d,Y^d)-m_{+,y}(\tilde{d}|D^d,Y^d)||_{R_{\min}^{-1}}^2 
		\end{align}
	\end{subequations}
	for all $(x,u,\check{w}), (\tilde{x}, u, \check{w}')$ with $x,\tilde{x} \in \mathbb{X}$ and $u \in \mathbb{U}$, where $d = \begin{bmatrix}
		x_1  &\dots& x_n &u_1 &\dots &u_m
	\end{bmatrix}^\top$, $\tilde{d} = \begin{bmatrix}
		\tilde{x}_1  &\dots& \tilde{x}_n &u_1 &\dots &u_m
	\end{bmatrix}^\top$, and $ Q_{\min}^{-1}$ and $R_{\min}^{-1}$ are from~(\ref{eq:Q}) and~(\ref{eq:R}), respectively.
\end{assumption}
Note that existence of a $\delta$-IOSS Lyapunov function is equivalent to the system being $\delta$-IOSS \cite{Allan2021}. This property is necessary and sufficient for the existence of state estimators for nonlinear systems and has widely been used in the recent MHE literature, compare \cite{Schiller2022} for a more detailed discussion.
\begin{remark}
	\label{rmk:GP_detectability}
	After having determined the posterior mean~(\ref{post_mean}) and variance~(\ref{post_var}), one can verify Assumption~\ref{ass:lyap} using the results of \cite[Sec. IV]{Schiller2022}. An interesting property for future research is to study whether Assumption~\ref{ass:lyap} is always satisfied (i.e., the learned GP model admits a $\delta$-IOSS Lyapunov function) if the true unknown system~(\ref{def:system}) admits a $\delta$-IOSS Lyapunov function (i.e., is detectable). 
\end{remark}

To simplify the notation in the following proof, we define  
\begin{align}
	\alpha_1^{\max} &\coloneqq \max_{\substack{x \in \mathbb{X}, u \in \mathbb{U}}} \Big\{\|f\big(x, u\big) - m_{+,x}\big(d|D^d,Y^d\big)\|_{Q_{\max}^{-1}} \Big\} \label{def:alpha1:max}  \\
	\alpha_2^{\max} &\coloneqq \max_{\substack{x \in \mathbb{X}, u \in \mathbb{U}}} \Big\{\|h\big(x, u\big) - m_{+,y}\big(d|D^d,Y^d\big)\|_{R_{\max}^{-1}}\Big\}  \label{def:alpha2:max}
\end{align}
and $ \alpha^{\max} \coloneqq \max \{\alpha_1^{\max}, \alpha_2^{\max}\}$. Notice that these constants exist, since we assume that (i) the states and the inputs evolve in compact sets, (ii) the functions $f$ and $h$ are continuous and since the here considered squared exponential ARD kernel leads to a continuous posterior mean.
\begin{theorem}
	\label{thm:MHE}
	Let Assumption~\ref{ass:lyap} hold. Then, there exist $\mu\in[0,1)$ and a minimal horizon length $\bar{M}$ such that for all $M\in\mathbb{I}_{\geq\bar{M}}$, the state estimation error of the GP based MHE (\ref{MHE_nom}) is bounded for all $t \in \mathbb{I}_{\geq 0}$ by
	\begin{align}
		\|\hat{x}&(t) - x(t)\|_{P_1} \leq \max \Bigg\{6\sqrt{\mu}^t\|\hat{x}(0) - x(0)\|_{P_2},\nonumber \\
		& \max_{q\in\mathbb{I}_{[0,t-1]}}\left\{\frac{12}{1-\sqrt[4]{\mu}}\sqrt[4]{\mu}^q\|  w(t-q-1) \|_{Q_{\max}^{-1}}\right\},  \nonumber \\ 
		& \max_{q\in\mathbb{I}_{[0,t-1]}} \left\{\frac{12}{1-\sqrt[4]{\mu}}\sqrt[4]{\mu}^q\|v(t-q-1)\|_{R_{\max}^{-1}}\right\},\nonumber \\
		& \hspace{1.5cm} \frac{12}{1-\sqrt[4]{\mu}} \alpha^{\max} \Bigg\}. \label{thm:eq:expression}
	\end{align}
	Consequently, the GP based MHE~(\ref{MHE_nom}) is pRES according to Definition~\ref{def:pRES}. 
\end{theorem}

The proof of Theorem~\ref{thm:MHE} can be found in Appendix~\ref{app:thm:proof}. It mainly relies on the Lyapunov-based robust stability proof technique recently proposed in the context of model-based MHE in \cite[Prop. 1, Thm. 1, Cor. 1]{Schiller2022}. Nevertheless, in our case we need to take into account two crucial differences, namely, (i) that the estimated system trajectory is based on the learned dynamics and (ii) that the weighting matrices in the cost function are not constant.

Theorem 1 shows that the state estimation error is upper bounded by means of (i) the initial state estimation error, (ii) the true process noise, (iii) the true measurement noise, and (iv) the mismatch between the learned system model and the true system dynamics. 
\begin{remark}
	\label{rmk:proof:counter-intuitive}
	In case of higher training data availability,~$R_{\min}^{-1}$ and $Q_{\min}^{-1}$ increase, since the maximal posterior variances decrease. Larger values of $R_{\min}^{-1}$ and $Q_{\min}^{-1}$ in~(\ref{ass:lyap:supply}) allow for a larger $W_{\delta}$, which then allows for a larger~$P_1$ in~(\ref{ass:lyap:bounds}). In turn, this results in a less conservative error bound~(\ref{thm:eq:expression}). Hence, a higher training data availability results in less conservative state estimation error bounds. Nevertheless, developing a proof technique that inherently makes use of the varying weighting matrices in order to reduce conservatism in the stability proof is an interesting subject for future research.
\end{remark}

Note that the estimation error bound~(\ref{thm:eq:expression}) of Theorem~\ref{thm:MHE} depends on $\alpha^{\max}$ that accounts for the mismatch between the learned system dynamics and the true system dynamics. A probabilistic upper bound for this mismatch can be obtained \cite{Lederer2019} by making the following additional assumption.
\begin{assumption}
	\label{ass:Lipschitz:continuity}
	Each component of the unknown functions~$f$ and~$h$ is Lipschitz continuous and a sample from a GP, i.e., $f_{i}$ is a sample of $\mathcal{GP}(0, k_{x_i}(d, d')),\: i = 1, \dots, n$ and $h_{j}$ a sample of $ \mathcal{GP}(0, k_{y_j}(d, d')),~j = 1, \dots, p.$
\end{assumption}

Furthermore, we define 
\begin{align}
	\beta(\tau) &\coloneqq 2 \log \bigg(\frac{B(\tau, \mathbb{X})}{\delta}\bigg) \label{eq:beta:cor}\\ 
	\gamma_{f_{1}}(\tau) &\coloneqq (L_{m_{+,x_1}}+L_{f_{1}}) \tau + \sqrt{\beta(\tau)} \omega_{\sigma_{+,x_1}}(\tau), \\ 
	\vdots \nonumber\\
    \gamma_{f_{n}}(\tau) &\coloneqq (L_{m_{+,x_n}}+L_{f_{n}}) \tau + \sqrt{\beta(\tau)} \omega_{\sigma_{+,x_n}}(\tau), \\
	\gamma_{h_{1}}(\tau) &\coloneqq (L_{m_{+,y_1}}+L_{h_{1}}) \tau + \sqrt{\beta(\tau)} \omega_{\sigma_{+,y_1}}(\tau), \\
	\vdots \nonumber\\
	\gamma_{h_{p}}(\tau) &\coloneqq (L_{m_{+,y_p}}+L_{h_{p}}) \tau + \sqrt{\beta(\tau)} \omega_{\sigma_{+,y_p}}(\tau) \label{eq:gamma:hp}, 
\end{align}
where $\tau$ corresponds to a grid constant. The constant $B$ denotes the covering number, which corresponds to the minimum number of points in a grid over $\mathbb{X}$ considering the grid constant~$\tau$. The constants $L_{m_{x_1}}, \dots, L_{m_{x_n}}, L_{m_{y_1}}, \dots, L_{m_{y_p}}$ denote the Lipschitz constants of the mean functions and $\omega_{\sigma_{+, x_1}}, \dots, \omega_{\sigma_{+, x_n}},\omega_{\sigma_{+, y_1}},\dots \omega_{\sigma_{+, y_p}}, $ the moduli of continuity of the kernels. Finally, $\delta \in (0,1)$ and $L_{f_{1}}, \dots, L_{f_{n}}, L_{h_{1}}, \dots, L_{h_{p}}$ are the Lipschitz constants of the components of the unknown functions of $f$ and $h$. The definitions (\ref{eq:beta:cor}) - (\ref{eq:gamma:hp}) were made in \cite{Lederer2019}, and the reader is referred to this reference for additional details. Moreover, we introduce 
\begin{align}
	\Delta^{\max}_x&(\tau) \coloneqq  \sqrt{\lambda_{\max}(Q_{\max}^{-1})} \times \nonumber \\
	 &\sum_{i = 1}^{n}\max_{x \in \mathbb{X}, u \in \mathbb{U}} \big\{ \| \sqrt{\beta(\tau)}\sigma_{+,x_i}(d|D^d,Y^d) + \gamma_{f_{i}}(\tau) 
	\| \big\}, \label{def:delta:max:x} \\ 
	\Delta^{\max}_y&(\tau) \coloneqq \sqrt{\lambda_{\max}(R_{\max}^{-1})} \times \nonumber \\ 
	&\sum_{i = 1}^{p} \max_{x \in \mathbb{X}, u \in \mathbb{U}} \big\{\| \sqrt{\beta(\tau)}\sigma_{+,y_i}(d|D^d,Y^d) + \gamma_{h_{i}}(\tau) 
	\| \big\}\label{def:delta:max:y} , 
\end{align}
which will be used in the following corollary to simplify the notation. 
\begin{corollary}
	\label{cor:prob_pRES}
	Let Assumptions~\ref{ass:lyap} -~\ref{ass:Lipschitz:continuity} hold. Then, there exist $\mu \in [0,1)$ and a minimal horizon length $\bar{M}$ such that for all $M \in \mathbb{I}_{\geq \bar{M}}$ the state estimation error of the GP based MHE~(\ref{MHE_nom}) is (probabilistically) bounded for all $t \in \mathbb{I}_{\geq 0}$ by
	\begin{align}
	P\Bigg(\|\hat{x}&(t) - x(t)\|_{P_1} \leq \max \Bigg\{6\sqrt{\mu}^t\|\hat{x}(0) - x(0)\|_{P_2},\nonumber \\
	& \max_{q\in\mathbb{I}_{[0,t-1]}}\left\{\frac{12}{1-\sqrt[4]{\mu}}\sqrt[4]{\mu}^q\|  w(t-q-1) \|_{Q_{\max}^{-1}}\right\},  \nonumber \\ 
	& \max_{q\in\mathbb{I}_{[0,t-1]}} \left\{\frac{12}{1-\sqrt[4]{\mu}}\sqrt[4]{\mu}^q\|v(t-q-1)\|_{R_{\max}^{-1}}\right\},\nonumber \\
	& \frac{12}{1-\sqrt[4]{\mu}}\Delta_x^{\max}(\tau), \frac{12}{1-\sqrt[4]{\mu}}\Delta_y^{\max}(\tau),\Bigg\}\Bigg) \nonumber\\
	&\geq (1-\delta)^{n+p}. \label{cor:math:expression}
\end{align}
\end{corollary}
The proof of Corollary~\ref{cor:prob_pRES} is shown in Appendix~\ref{app:cor:proof}. The key idea of the proof is to bound each component of the difference between the functions~$f$,~$h$ and the posterior means by applying the probabilistic bound developed in \cite[Thm. 3.1]{Lederer2019}.

Corollary~\ref{cor:prob_pRES} uses a probabilistic upper bound for the mismatch between the learned and the true unknown dynamics. As a result, also the obtained estimation error bound (\ref{cor:math:expression}) is probabilistic in nature. Note that the final probability in (\ref{cor:math:expression}) decreases for a higher state/output dimension. This is due to the current (conservative) proof technique of bounding component-wise the difference between the functions $f$, $h$ and the posterior means. Developing a less conservative proof is an interesting subject for future research.

As can be seen from the definition of~$\beta$ in (\ref{eq:beta:cor}), a higher probability (i.e., a smaller~$\delta$) results in more conservative upper bounds and vice versa. Furthermore, as discussed in Remark~\ref{rmk:proof:counter-intuitive}, more training data will improve the posterior variance and thus the (probabilistic) estimation error bounds (\ref{cor:math:expression}) that explicitly depend on $\sigma_{+,x_1}, \dots,~\sigma_{+,x_n}$ and $\sigma_{+,y_1},\dots,~\sigma_{+,y_p}$.

\section{APPLICATION TO BATCH REACTOR SYSTEM}
\label{sec:Numerical-Example}
In this section, we illustrate the performance of the GP based MHE. To this end, we consider the following Euler-discretized system 
\begin{align*}
	x_1(t+1) &= x_1(t) + T(-2k_1 x_1^2(t) + 2k_2x_2(t)) + w_1(t)\\
	x_2(t+1) &= x_2(t) + T(k_1x_1^2(t) - k_2x_2(t)) + w_2(t) \\
	y(t) &= x_1(t) + x_2(t) + v(t) 
\end{align*}
with sampling time $T = 0.1$, constants $k_1 = 0.16$, $k_2 = 0.0064$ which corresponds to a batch reactor system \cite[Ch. 4]{Rawlings2020}, \cite{Tenny2002}. This system is a benchmark example in the MHE literature, since other nonlinear state estimation techniques, such as the extended Kalman filter, can fail to converge, compare, \cite{Rawlings2020}. 

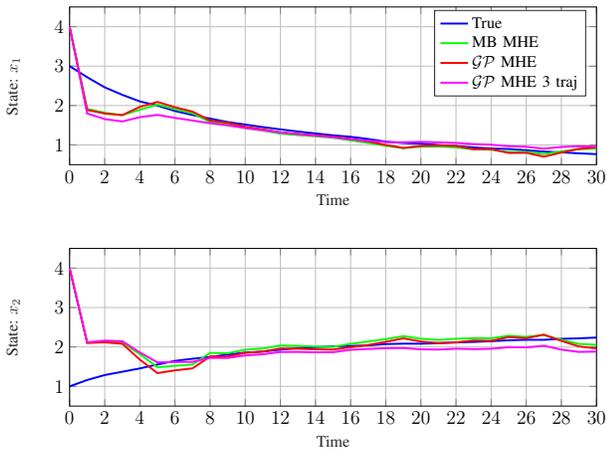
\begin{figure}[t!]
	\centering
	\scalebox{0.61}{
%
%

\definecolor{mycolor1}{rgb}{1.00000,0.00000,1.00000}%
\begin{tikzpicture}

\begin{axis}[%
width=4.521in,
height=1.354in,
at={(0.758in,2.693in)},
scale only axis,
xmin=0,
xmax=30,
xlabel style={font=\color{white!15!black}},
xlabel={Time},
ymin=0.5,
ymax=4.5,
ylabel style={font=\color{white!15!black}},
ylabel={State: $x_1$},
axis background/.style={fill=white},
xmajorgrids,
ymajorgrids,
legend style={legend cell align=left, align=left, draw=white!15!black}
]
\addplot [color=blue, line width=1.2pt]
  table[row sep=crcr]{%
0	3\\
1	2.71865667139546\\
2	2.46104016621207\\
3	2.27206124851007\\
4	2.10428761062838\\
5	2.00023839909907\\
6	1.86069761704114\\
7	1.75926951747608\\
8	1.66955287855036\\
9	1.58135065400919\\
10	1.51773052242434\\
11	1.45311213851699\\
12	1.39512444095257\\
13	1.34020106984708\\
14	1.29251660186043\\
15	1.24455055502055\\
16	1.20644643028141\\
17	1.15177361355314\\
18	1.08249022506746\\
19	1.05089900863837\\
20	1.03192819228345\\
21	0.999495383724181\\
22	0.973402391843299\\
23	0.93713860220596\\
24	0.910109181101536\\
25	0.897283246359676\\
26	0.865659338818742\\
27	0.832330527446711\\
28	0.812884643997367\\
29	0.786866379088311\\
30	0.767637336639605\\
};
\addlegendentry{True}

\addplot [color=green, line width=1.2pt]
  table[row sep=crcr]{%
0	4\\
1	1.91415278422196\\
2	1.81751103297809\\
3	1.75709038969973\\
4	1.89417649997059\\
5	2.02286434596217\\
6	1.90949593615199\\
7	1.8072112391318\\
8	1.57890012725275\\
9	1.52708994567675\\
10	1.42813140198154\\
11	1.36603756942707\\
12	1.288416006402\\
13	1.25160029073591\\
14	1.21957537427769\\
15	1.18392910224947\\
16	1.11814055500959\\
17	1.05170151615586\\
18	0.979560404879422\\
19	0.918165209499168\\
20	0.955208260847827\\
21	0.95862490557839\\
22	0.939068548583429\\
23	0.887736366699776\\
24	0.881855560773871\\
25	0.820781672897281\\
26	0.824088657561574\\
27	0.769619105872804\\
28	0.836162573728626\\
29	0.892968171304067\\
30	0.907490544984791\\
};
\addlegendentry{MB MHE}

\addplot [color=red, line width=1.2pt]
  table[row sep=crcr]{%
0	4\\
1	1.88742093312753\\
2	1.79799510014402\\
3	1.75665612312494\\
4	1.96690181410694\\
5	2.09159877026788\\
6	1.95759956461874\\
7	1.84676466439142\\
8	1.61891044771071\\
9	1.56200434978584\\
10	1.46120019238483\\
11	1.39706438725367\\
12	1.31880051190678\\
13	1.28004959642542\\
14	1.24647138846643\\
15	1.20981291507332\\
16	1.15474599406612\\
17	1.08767438869683\\
18	1.00005054924209\\
19	0.925057370250883\\
20	0.974503632439521\\
21	0.984668323704093\\
22	0.972777467867706\\
23	0.892965164059435\\
24	0.891152549688401\\
25	0.797910819943484\\
26	0.800776399089497\\
27	0.702564006633034\\
28	0.807255728740509\\
29	0.901085406802739\\
30	0.938857817789083\\
};
\addlegendentry{$\mathcal{GP}$ MHE}

\addplot [color=mycolor1, line width=1.2pt]
  table[row sep=crcr]{%
0	4\\
1	1.79692885478395\\
2	1.65671635625696\\
3	1.59617571261685\\
4	1.70483330922882\\
5	1.76292851945643\\
6	1.68678296485863\\
7	1.61966447239785\\
8	1.55184062471156\\
9	1.49671944770808\\
10	1.4291636181245\\
11	1.37317275037755\\
12	1.31543058691637\\
13	1.27357687569544\\
14	1.2358942808856\\
15	1.1957364205407\\
16	1.1418335633007\\
17	1.10728369512385\\
18	1.07112696392804\\
19	1.06944219059771\\
20	1.07858001740872\\
21	1.06486622251562\\
22	1.05179434641114\\
23	1.02190169645442\\
24	1.00692482589858\\
25	0.968813120819771\\
26	0.954056432105951\\
27	0.905991541498593\\
28	0.949306178138687\\
29	0.967491964081282\\
30	0.957377201843882\\
};
\addlegendentry{$\mathcal{GP}$ MHE 3 traj}

\end{axis}

\begin{axis}[%
width=4.521in,
height=1.354in,
at={(0.758in,0.62in)},
scale only axis,
xmin=0,
xmax=30,
xlabel style={font=\color{white!15!black}},
xlabel={Time},
ymin=0.5,
ymax=4.5,
ylabel style={font=\color{white!15!black}},
ylabel={State: $x_2$},
axis background/.style={fill=white},
xmajorgrids,
ymajorgrids
]
\addplot [color=blue, line width=1.2pt, forget plot]
  table[row sep=crcr]{%
0	1\\
1	1.16169885014595\\
2	1.28783460163631\\
3	1.37084100372355\\
4	1.45598610721826\\
5	1.55359706798039\\
6	1.64696725897189\\
7	1.70067778114725\\
8	1.74706015454625\\
9	1.8054375211536\\
10	1.85846468352933\\
11	1.88205649192419\\
12	1.9309388868547\\
13	1.97119197135818\\
14	1.99563422177056\\
15	2.01321377448839\\
16	2.02523711394706\\
17	2.03913418307611\\
18	2.07343825943713\\
19	2.08331053715725\\
20	2.08253227384665\\
21	2.09582299547978\\
22	2.11359411108243\\
23	2.12710109335251\\
24	2.14606848168397\\
25	2.16904051037624\\
26	2.18130779094806\\
27	2.18076660400114\\
28	2.20578160216814\\
29	2.2186561931815\\
30	2.23831635378264\\
};
\addplot [color=green, line width=1.2pt, forget plot]
  table[row sep=crcr]{%
0	4\\
1	2.11104366226742\\
2	2.13033678953143\\
3	2.11652113381389\\
4	1.81792638064631\\
5	1.48508079302157\\
6	1.5215611091751\\
7	1.55499990339001\\
8	1.85133303483994\\
9	1.83933764523264\\
10	1.93157839844688\\
11	1.96146254606703\\
12	2.0370539930713\\
13	2.02878813291714\\
14	2.01012524762867\\
15	2.00361815309252\\
16	2.07868171325013\\
17	2.13749743642911\\
18	2.19885472335072\\
19	2.2763832129187\\
20	2.20808743391927\\
21	2.17943484866449\\
22	2.20741370290648\\
23	2.22347813970916\\
24	2.21962765040915\\
25	2.2841862983946\\
26	2.25528857470219\\
27	2.30366835928256\\
28	2.18059225681923\\
29	2.07790158929931\\
30	2.0539134577763\\
};
\addplot [color=red, line width=1.2pt, forget plot]
  table[row sep=crcr]{%
0	4\\
1	2.09991384873931\\
2	2.11768017093581\\
3	2.08184238915881\\
4	1.67812111906641\\
5	1.33701537622723\\
6	1.40658852766249\\
7	1.45711101235035\\
8	1.75731562927822\\
9	1.7542092378939\\
10	1.85061392062124\\
11	1.88427599762146\\
12	1.96086326907603\\
13	1.95549972185362\\
14	1.93933646752285\\
15	1.93424147018208\\
16	1.99311617429379\\
17	2.05104570207913\\
18	2.12871838038633\\
19	2.21912282423649\\
20	2.13731074764539\\
21	2.10027089583867\\
22	2.11757619137332\\
23	2.16379735078907\\
24	2.15438375482479\\
25	2.25073365814199\\
26	2.2218249825244\\
27	2.31287873330305\\
28	2.1529560775646\\
29	2.0136012063265\\
30	1.96381053963042\\
};
\addplot [color=mycolor1, line width=1.2pt, forget plot]
  table[row sep=crcr]{%
0	4\\
1	2.11891792553781\\
2	2.1608802851227\\
3	2.14595924682827\\
4	1.8613211557117\\
5	1.60865753121993\\
6	1.61716821150643\\
7	1.62206324065296\\
8	1.72057783796946\\
9	1.72153173867863\\
10	1.78464578970975\\
11	1.81475787832969\\
12	1.87259350042949\\
13	1.87435313436457\\
14	1.86511725306268\\
15	1.86754329915366\\
16	1.9274460021799\\
17	1.94902157928899\\
18	1.96820303310912\\
19	1.97371278975623\\
20	1.94016224509917\\
21	1.9341178507559\\
22	1.95585450557317\\
23	1.94523302609486\\
24	1.95428532860674\\
25	1.99419714934656\\
26	1.98769537875375\\
27	2.0317588653093\\
28	1.9318419074446\\
29	1.87444344457575\\
30	1.88217948091179\\
};
\end{axis}
\end{tikzpicture}%
}
	\caption{Simulation results of the GP based MHE scheme~(\ref{MHE_nom}).}
	\label{fig:x1}
\end{figure}

As mentioned in Section~\ref{sec:pre}, we consider two different phases. In both phases, we consider $w \sim \mathcal{N}(0, \sigma_w^2I_n)$ with $\sigma_w = 0.01$ and $v \sim \mathcal{N}(0, \sigma_v^2I_p)$ with $\sigma_v = 0.1$. In the offline phase, we collect five different state/output trajectories (of length 31) with the following initial conditions $x_{01} = \begin{bmatrix}
		3 & 1
\end{bmatrix}^\top$, $x_{02} =\begin{bmatrix}
	 1.2 & 4.5
\end{bmatrix}^\top$, $x_{03} =\begin{bmatrix}
0.5 & 3.5
\end{bmatrix}^\top $, $x_{04} = \begin{bmatrix}
1 & 3
\end{bmatrix}^\top$, $x_{05} = \begin{bmatrix}
2 & 4
\end{bmatrix}^\top$ and perform the hyperparameter optimization by maximizing the log marginal likelihood. In the online phase, we apply the MHE scheme~(\ref{MHE_nom}) using $\eta = 0.91$, $M = 15$, initial condition $x_0 =\begin{bmatrix}
3 & 1
\end{bmatrix}^\top$, and\footnote{
In \cite[Cor. 3]{Schiller2022}, a method to verify the existence of the $\delta$-IOSS Lyapunov function and choose the values of $P_2$ and $\eta$ was proposed. However, the application of this result to the proposed GP based scheme is not straightforward (since the resulting posterior mean function is highly complex) and an interesting subject for future research.
} $P_2 = I_n$, $R_0 = 100$ and $Q_0 = \mathrm{diag}(1000, 1000)$. As in \cite[Sec. V]{Schiller2022}, we consider that the states evolve in a compact set $\mathbb{X} = \{x\in \mathbb{R}^2: 0.1 \leq x_i \leq 4.5, \: i =\{1,2\}\}$. In addition, we illustrate the performance of the GP based MHE with the same parameters, when only three trajectories (for initial conditions $x_{01} =\begin{bmatrix}
0.5 & 3.5
\end{bmatrix}^\top $, $x_{04} = \begin{bmatrix}
1 & 3
\end{bmatrix}^\top$, $x_{05} = \begin{bmatrix}
2 & 4
\end{bmatrix}^\top$), i.e., less data, are collected in the offline phase. Finally, we implement a standard model-based MHE scheme (that is based on exact model knowledge) with the same characteristics, i.e., using the same $P_2$, $Q_0$, $R_0$, $\eta$, and $M$. The obtained results are illustrated in Figure~\ref{fig:x1}. As guaranteed by Theorem~\ref{thm:MHE}, the GP based MHE scheme is robustly stable. The estimation performance improves when more training data is available. Furthermore, the GP based MHE (related to five collected trajectories) performs similarly well as the model-based MHE.

\section{CONCLUSION}
\label{sec:Conclusion}
In this paper, we introduced a GP based MHE framework for which we proved practical robust exponential stability. The framework leverages the posterior mean of the GP to replace the required mathematical model in the MHE scheme and the posterior variance to account for the uncertainty of the learned model within the cost function. This allows for an effective way to estimate the states of unknown nonlinear systems, as was also illustrated by a batch reactor example.

Future work includes investigating how detectability ($\delta$-IOSS) can efficiently be verified for the learned system, in particular whether/when the learned model inherits this property if the true unknown system is detectable.

\bibliographystyle{IEEEtran}
\bibliography{IEEEabrv,cdc2023}
\appendix
\subsection{Proof of Theorem~\ref{thm:MHE}}
\label{app:thm:proof}
The proof of Theorem~\ref{thm:MHE} is based on the developments shown in \cite{Schiller2022}. Here, we mainly comment on the steps that are conceptually different from the proof in \cite{Schiller2022}, without describing the similar steps of the proof in all detail.

\textbf{Proof:}
The constraints in the MHE problem guarantee that the (optimal) estimated system trajectory (denoted by~$\hat{x}(j|t), u(t), \hat{w}(j|t), \hat{v}(j|t)$ for all $j \in \mathbb{I}_{[t- M_t, t-1]}$) fulfills the learned system dynamics. The unknown (true) system trajectory cannot necessarily be represented by the posterior means. Therefore, we use the introduced auxiliary variables~$\check{w}$~(\ref{def:auxiliary_w}) and $\check{v}$~(\ref{def:auxiliary_v}) to represent the true system trajectory by the posterior means in the following. Exploiting that due to~(\ref{eq:mean_in_MHE_h}) $m_{+,y}(d(t-j)|D^d,Y^d) - m_{+,y}(\hat{d}(t-j|t)|D^d,Y^d) = \hat{v}(t-j|t)-\check{v}(t-j)$ for all $j\in\mathbb{I}_{[t-M_t,t-1]}$ and applying inequality (\ref{ass:lyap:supply}) $M_t$ times yields
\begin{align*}
	W_\delta (\hat{x}(t), x(t)) \overset{(\ref{ass:lyap:supply})}{\leq}& \sum_{j =1 }^{M_t} \eta^{j-1} \Big(||\hat{w}(t-j|t) - \check{w}(t-j)||_{Q_{\min}^{-1}}^2 \nonumber \\
	&+ ||\hat{v}(t-j|t) - \check{v}(t-j)||_{R_{\min}^{-1}}^2 \Big) \nonumber \\
	&+ \eta^{M_t}  W_\delta(\hat{x}(t-M_t|t), x(t-M_t)).
\end{align*}
Moreover, using (\ref{ass:lyap:bounds}), the triangle inequality, the Cauchy-Schwarz inequality, and Young's inequality, it holds
\begin{align*}
	W_\delta (\hat{x}(t)&, x(t)) \\ 
	\leq&  \sum_{j =1 }^{M_t} \eta^{j-1} \Big(2||\hat{w}(t-j|t)||_{Q_{\min}^{-1}}^2 + 2||\check{w}(t-j)||_{Q_{\min}^{-1}}^2 \\
	&+ 2||\hat{v}(t-j|t)||^2_{R_{\min}^{-1}} +2||\check{v}(t-j)||_{R_{\min}^{-1}}^2 \Big) \nonumber \\
	&+ 2\eta^{M_t}  ||\hat{x}(t-M_t) - x(t-M_t) ||_{P_2}^2 \\
	&+ 2\eta^{M_t}||\hat{x}(t-M_t|t) -  \hat{x}(t-M_t)||_{P_2}^2.
\end{align*}
We rearrange the terms and obtain
\begin{align}
	W_\delta &(\hat{x}(t), x(t)) \nonumber\\
	\leq&2\eta^{M_t} 
	||\hat{x}(t-M_t) - x(t-M_t) ||_{P_2}^2 \nonumber\\
	&+\sum_{j =1 }^{M_t} 2\eta^{j-1}\Big(||\check{w}(t-j)||_{Q_{\min}^{-1}}^2  + ||\check{v}(t-j)||_{R_{\min}^{-1}}^2 \Big) \nonumber \\
	&+ 2 \eta^{M_t} ||\hat{x}(t-M_t|t) -  \hat{x}(t-M_t)||_{P_2}^2 \nonumber\\
	&+  \sum_{j = 1}^{M_t} 2\eta^{j-1}\Big(||\hat{w}(t-j|t)||_{Q_{\min}^{-1}}^2  + ||\hat{v}(t-j|t)||_{R_{\min}^{-1}}^2 \Big).\nonumber \\ 
	\eqqcolon& 2\eta^{M_t}||\hat{x}(t-M_t) - x(t-M_t) ||_{P_2}^2 \nonumber \\ 
	&+\sum_{j =1 }^{M_t} 2\eta^{j-1}\Big(||\check{w}(t-j)||_{Q_{\min}^{-1}}^2  + ||\check{v}(t-j)||_{R_{\min}^{-1}}^2\Big) \nonumber \\
	&+ J_{\min}(\hat{x}(t-M_t|t), \hat{w}(\cdot|t), \hat{v}(\cdot|t),t). \label{Proposition1_previous}
\end{align}
Note that $J_{\min}(\hat{x}(t-M_t|t), \hat{w}(\cdot|t), \hat{v}(\cdot|t),t)$ does not correspond to the optimal cost of problem (\ref{MHE_nom}). In fact, $J_{\min}(\hat{x}(t-M_t|t), \hat{w}(\cdot|t), \hat{v}(\cdot|t),t)$ corresponds to the cost of the optimal trajectory, when $R_{\min}^{-1}$, $Q_{\min}^{-1}$ (and $P_2$) are considered in the cost function (\ref{eq:cost:function}) (but \textit{not} the variable $R^{-1}_{\bar{d}(t-j|t)}$, $Q^{-1}_{\bar{d}(t-j|t)}$). Analogously, we define $J_{\max}$ by replacing $R^{-1}_{\bar{d}(t-j|t)}$ and $Q^{-1}_{\bar{d}(t-j|t)}$ in (\ref{eq:cost:function}) with $R_{\max}^{-1}$ and  $Q_{\max}^{-1}$, respectively. Next, we upper bound $J_{\min}(\hat{x}(t-M_t|t),\hat{w}(\cdot|t),\hat{v}(\cdot|t),t)$ as follows
\begin{align*}
	J_{\min}(\hat{x}(t-M_t|t),& \hat{w}(\cdot|t), \hat{v}(\cdot|t),t) \\
	&\leq 			J^\ast(\hat{x}(t-M_t|t), \hat{w}(\cdot|t), \hat{v}(\cdot|t),t) \\
	&\leq J(x(t-M_t), \check{w}(\cdot), \check{v}(\cdot),t) \\
	&\leq J_{\max}(x(t-M_t), \check{w}(\cdot), \check{v}(\cdot),t),
\end{align*}	
where the first inequality holds by (\ref{eq:Q}) - (\ref{eq:R}), the second is due to optimality (i.e., the true unknown system trajectory $x(\cdot), \check{w}(\cdot), \check{v}(\cdot)$ is a feasible but in general suboptimal solution to problem (\ref{MHE_nom})), and the third again follows from (\ref{eq:Q}) - (\ref{eq:R}). We consider these bounds in inequality (\ref{Proposition1_previous}) and obtain
\begin{align*}
	W_\delta &(\hat{x}(t), x(t)) \\
	\leq& 2\eta^{M_t}||\hat{x}(t-M_t) - x(t-M_t) ||_{P_2}^2 \\
	&+\sum_{j =1 }^{M_t} 2\eta^{j-1}\Big(||\check{w}(t-j)||_{Q_{\min}^{-1}}^2  + ||\check{v}(t-j)||_{R_{\min}^{-1}}^2\Big)\nonumber \\
	&+ J_{\max}(x(t-M_t|t), \check{w}(\cdot), \check{v}(\cdot),t).
\end{align*}
Using~(\ref{eq:Q}) and~(\ref{eq:R}), we obtain
\begin{align}
	W_\delta &(\hat{x}(t), x(t)) \nonumber\\
	\leq &   4\eta^{M_t}||\hat{x}(t-M_t) - x(t-M_t) ||_{P_2}^2 \nonumber\\
	&+\sum_{j =1 }^{M_t} 4\eta^{j-1}\Big(||\check{w}(t-j)||_{Q_{\max}^{-1}}^2  + ||\check{v}(t-j)||_{R_{\max}^{-1}}^2\Big) \label{eq:afterJmax}\\
	\leq& 4\lambda_{\max}(P_2, P_1)\eta^{M_t}
	W_\delta (\hat{x}(t-M_t),x(t-M_t)) \nonumber\\
	&+\sum_{j =1 }^{M_t} 4\eta^{j-1}\Big(||\check{w}(t-j)||_{Q_{\max}^{-1}}^2  + ||\check{v}(t-j)||_{R_{\max}^{-1}}^2\Big). \nonumber
\end{align}
We choose $M$ large enough such that $\mu^{M} \coloneqq 4\lambda_{\max}(P_2, P_1)\eta^{M} <1$ with $\mu \in [0,1)$, and obtain for all $t\in\mathbb{I}_{\geq M}$
\begin{align}
	W_\delta &(\hat{x}(t), x(t)) \leq \mu^{M} W_\delta (\hat{x}(t-M),x(t-M)) \nonumber \\
	&+\sum_{j =1 }^{M} 4\eta^{j-1}\Big(||\check{w}(t-j)||_{Q_{\max}^{-1}}^2  + ||\check{v}(t-j)||_{R_{\max}^{-1}}^2\Big) \label{eq:M_step}.
\end{align}
Performing similar steps as the ones in the proof of \cite[Cor. 1]{Schiller2022} results in the following state estimation error bound
\begin{align}
	\|\hat{x}(t)& - x(t)\|_{P_1} \nonumber  \\
	&\leq\ \max \Bigg\{ 6\sqrt{\mu}^t\|\hat{x}(0) - 	x(0)\|_{P_2}, \nonumber\\
	&\max_{q\in\mathbb{I}_{[0,t-1]}}\left\{\frac{6}{1-\sqrt[4]{\mu}}\sqrt[4]{\mu}^q\|\check{w}(t-q-1)\|_{Q_{\max}^{-1}}\right\},\nonumber \\ 
	&\max_{q\in\mathbb{I}_{[0,t-1]}}\left\{\frac{6}{1-\sqrt[4]{\mu}}\sqrt[4]{\mu}^q\|\check{v}(t-q-1)\|_{R_{\max}^{-1}}\right\} \Bigg\}. \label{final_result}
\end{align}
We replace $\check{w}$ and $\check{v}$ according to~(\ref{def:auxiliary_w}) and~(\ref{def:auxiliary_v}), respectively. Then, we apply the triangle inequality and bound the difference between the functions $f$, $h$ and the posterior means $m_{+,x}$, $m_{+,y}$ by~(\ref{def:alpha1:max}) and~(\ref{def:alpha2:max}), respectively, which results in 
\begin{align}
	\|&\hat{x}(t) - x(t)\|_{P_1} \leq \max \Bigg\{ 6\sqrt{\mu}^t\|\hat{x}(0) - 	x(0)\|_{P_2}, \nonumber \\ 
	&\max_{q\in\mathbb{I}_{[0,t-1]}}\left\{\frac{6}{1-\sqrt[4]{\mu}}\sqrt[4]{\mu}^q(\|w(t-q-1)\|_{Q_{\max}^{-1}} + \alpha_1^{\max})\right\}, \nonumber \\
	&\max_{q\in\mathbb{I}_{[0,t-1]}}\left\{\frac{6}{1-\sqrt[4]{\mu}}\sqrt[4]{\mu}^q(\|v(t-q-1)\|_{R_{\max}^{-1}}  + \alpha_2^{\max})\right\} \Bigg\}. 
\end{align}
Using that $\max_{q\in\mathbb{I}_{[0,t-1]}}\sqrt[4]{\mu}^q =1$ and $a+b \leq \max\{2a,2b\}$ for any $a,b \geq 0$, we have
\begin{align}
	\|\hat{x}(t) &- x(t)\|_{P_1} \leq \max \Bigg\{ 6\sqrt{\mu}^t\|\hat{x}(0) - 	x(0)\|_{P_2}, \nonumber \\ 
	&\max_{q\in\mathbb{I}_{[0,t-1]}}\left\{\frac{12}{1-\sqrt[4]{\mu}}\sqrt[4]{\mu}^q\|w(t-q-1)\|_{Q_{\max}^{-1}} \right\},\nonumber \\
	&\max_{q\in\mathbb{I}_{[0,t-1]}}\left\{\frac{12}{1-\sqrt[4]{\mu}}\sqrt[4]{\mu}^q\|v(t-q-1)\|_{R_{\max}^{-1}} \right\}, \nonumber \\ 
	&\hspace{1cm} \frac{12}{1-\sqrt[4]{\mu}} \alpha_1^{\max},\frac{12}{1-\sqrt[4]{\mu}}\alpha_2^{\max}\Bigg\} \label{eq:nearly:final:result}. 
\end{align}
Finally, we use $\alpha_{\max}$ to bound the last two terms of~(\ref{eq:nearly:final:result}), which leads to the expression of Theorem~\ref{thm:MHE}. \hfill$\blacksquare$

\subsection{Proof of Corollary~\ref{cor:prob_pRES}}
\label{app:cor:proof}
\textbf{Proof:} 
From the expressions in (\ref{def:alpha1:max}) - (\ref{def:alpha2:max}), we can bound $\alpha_1^{\max}$ and $\alpha_2^{\max}$ as follows 
	\begin{align}
		\alpha_1^{\max} &\leq \sqrt{\lambda_{\max}(Q_{\max}^{-1})}  \times \nonumber \\
		&\sum_{i = 1}^{n} \max_{x \in \mathbb{X}, u \in \mathbb{U}} \big\{ || f_{i}(x,u) -m_{+, x_i}(d|D^d,Y^d)|| \big \}. \label{eq:alpha:max1:prob} \\
		\alpha_2^{\max} &\leq \sqrt{\lambda_{\max}(R_{\max}^{-1})} \times \nonumber  \\ 
		&\sum_{i = 1}^{p} \max_{x \in \mathbb{X}, u \in \mathbb{U}} \big\{|| h_{i}(x,u) -m_{+, y_i}(d|D^d,Y^d)|| \big\}. \label{eq:alpha:max2:prob}
	\end{align} 
	From here on, we apply \cite[Thm 3.1]{Lederer2019} to probabilistically bound the difference between the true function components of $f$, $h$ and the corresponding posterior means, which results in
	\begin{align*}
		&P\Big(||f_{i}(x,u) - m_{x_i}(d|D^d,Y^d)||\\
		& \hspace{0.5cm} \leq \sqrt{\beta(\tau)}\sigma_{+, x_i}(d|D^d,Y^d) + \gamma_{f_{i}}(\tau), \forall x \in \mathbb{X}, u \in \mathbb{U}\Big) \\ 
		&\hspace{2cm}\geq 1- \delta  \hspace{2cm} i = 1, \dots, n \\
		&P\Big(||h_{j}(x,u) - m_{y_j}(d|D^d,Y^d)|| \\ 
		& \hspace{0.5cm} \leq \sqrt{\beta(\tau)}\sigma_{+, y_j}(d|D^d,Y^d) + \gamma_{h_{j}}(\tau), \forall x \in \mathbb{X}, u \in \mathbb{U}\Big) \\
		&\hspace{2cm}\geq 1- \delta  \hspace{2cm} j = 1, \dots, p.	 
	\end{align*}
	 Since the GPs are considered to be independent, the probability that all the components of $f$ jointly fulfill their bounds is lower bounded by by $(1-\delta)^n$ (the same holds for $h$ with probability $(1-\delta)^p$). We replace the right-hand sides of (\ref{eq:alpha:max1:prob}) and~(\ref{eq:alpha:max2:prob}) by these probabilistic bounds, i.e., 
	 \begin{align}
 		P\big(\alpha_1^{\max} &\leq \Delta_x^{\max}(\tau)\big) \geq (1- \delta)^n \\
	 	P\big(\alpha_2^{\max} &\leq \Delta_y^{\max}(\tau)\big) \geq (1-\delta)^p.
	 \end{align}
	 Finally, the probability that both $\alpha_1^{\max} \leq \Delta_x^{\max}(\tau)$ and $\alpha_2^{\max} \leq \Delta_y^{\max}(\tau)$ hold jointly is lower bounded by $(1-\delta)^{(n+p)}$. Using this bound in (\ref{eq:nearly:final:result}), we obtain the left-hand side of~(\ref{cor:math:expression}). \hfill $\blacksquare$
\end{document}